\documentclass[11pt,preprint]{aastex}

\newcommand{\mtwo}{IRAS~00117+6412}
\newcommand{\mten}{IRAS~05274+3345}
\newcommand{\mfourteen}{IRAS~05553+1631}
\newcommand{\mfifteen}{IRAS~06056+2131}
\newcommand{\msixteen}{IRAS~06061+2151}
\newcommand{\mnineteen}{IRAS~06103+1523}
\newcommand{\mtwoone}{IRAS~06105+1756}
\newcommand{\mtwoseven}{IRAS~06382+0939}
\newcommand{\monetwofour}{IRAS~20227+4154}
\newcommand{\monefoursix}{IRAS~22267+6244}
\newcommand{\monefourseven}{IRAS~22272+6358A}
\newcommand{\monesixthree}{IRAS~23545+6508}

\newcommand{\kms}{km~s$^{-1}$}

\newcommand\coj{CO J=2$-$1}

\newcommand{\msol}{$M_\odot$}
\newcommand{\lsol}{$L_\odot$}

\newcommand{\tr}{$T_{\rm R}^*$}
\newcommand{\tb}{$T_{\rm b}$}

\begin{document}


\title{
Occurrence Frequency of CO Outflows in Massive Protostellar Candidates 
}

\author{Kee-Tae Kim\altaffilmark{1,2} and S. E. Kurtz\altaffilmark{3}}
\altaffiltext{1}{Department of Astronomy, University of Illinois,
1002 West Green Street, Urbana, IL 61801}
\altaffiltext{2}{Korea Astronomy \& Space Science Institute,
61-1 Hwaam-Dong, Yuseong-Gu, Daejeon 305-348, Korea; ktkim@kasi.re.kr}
\altaffiltext{3}{Centro de Radioastronom\'\i{}a y Astrof\'\i{}sica, UNAM, 
Apartado Postal 3-72, Morelia 58089, Michoac\'an, M\'exico; s.kurtz@astrosmo.unam.mx}



\begin{abstract}

We mapped 12 massive protostellar candidates in the \coj\ line,
which in combination with Zhang et al. (2005) completes
an unbiased survey of outflows for all 48 sources with $l$$>$50\degr\
in a sample of 101 massive protostellar candidates.
We detected outflows in 10 sources, implying 88\% occurrence frequency 
of outflows for the 48 sources. 
This supports the conclusion of previous studies that bipolar outflows are 
an integral component
in the formation process of massive stars.
The vast majority of the observed outflows are much more massive ($>$10~\msol)
and energetic ($>$100~\msol~\kms) than outflows from low-mass protostars.
They also have large mass outflow rates 
($>$2$\times$10$^{-4}$~\msol~yr$^{-1}$), 
suggesting large ($\sim$1$\times$10$^{-4}$~\msol~yr$^{-1}$) accretion rates
sufficient to overcome radiation pressure of the central massive protostars.
We compared the frequency distribution of collimation factors of 
40 massive outflows including those of this study 
with that of 36 low-mass outflows from the literature, 
and found {\it no} significant difference between the two.
All these results are consistent with the suggestion that massive stars
form through accretion as do low-mass stars but with much higher 
accretion rates.

\end{abstract}





\section{Introduction}

It has been well established that low-mass stars form
by gravitational collapse of dense molecular cores and that
accretion disks and bipolar outflows are basic building blocks
of the process \citep{shu87}.
Compared to low-mass star formation, very little is known about
the formation process of massive ($>$8~M$_\odot$) stars,
such that it is still much debated whether  massive stars form in
a qualitatively similar manner to low-mass stars.
This debate is mainly because the radiation pressure of massive stars
is strong enough to reverse materials infalling at the typical 
mass accretion rates observed in low-mass young stellar objects (YSOs)
\citep{wc87}.
There are currently two major theoretical models
in competition: accretion via disks \citep[e.g.,][]{ys02, mt03}
and coalescence of low-mass (proto)stars \citep[e.g.,][]{bon98}.
The most convincing way to differentiate between the two models 
may be to investigate whether well-defined accretion disks exist around
massive YSOs. 
But it is difficult to detect the accretion disks, if any,
because they would be small (at most several hundred AU),
short-lived, and easily confused by envelopes.  
In this study we attempt to distinguish between the two models
by observing bipolar molecular outflows that are much easier to detect.

Bipolar outflows from low-mass YSOs have been extensively
studied since the early 1980's 
\citep[see][and references therein]{bac96},
while systematic studies of outflows associated with
massive star formation began much later.
These studies are roughly divided into two groups.
The first group are single-point CO line surveys towards samples of 
massive YSOs in search of high-velocity molecular gas  
\citep[e.g,][]{sc96a,sri02}.
They showed that high-velocity gas is a common feature in massive YSOs.
\citet{sc96a} made a CO J=1$-$0
line survey towards 122 massive YSOs, the vast majority of which are
ultracompact (UC) HII regions, and detected high-velocity gas in 
90\% of the 94 sources suitable for line wing analysis. 
\citet{sri02} obtained a similar detection rate (84\%) from 
a CO J=2$-$1 line survey of 69 massive protostellar candidates.
The second group are CO line mappings of subsets of the sources that exhibit 
high-velocity wings 
\citep[e.g.,][]{sc96b,beu02a}.
These studies investigated whether the high-velocity wings are 
caused by bipolar outflows
and derived the physical and dynamical parameters of the outflows. 
\citet{sc96b} mapped 10 sources with high-velocity gas 
and identified outflows in 5 sources. If we assume a similar detection rate
for the remainder of their sample, the occurrence frequency of
outflows would be about 45\% for their sample.
\citet{beu02a} found outflows in 21 of 26 sources showing 
high-velocity wings in the sample of \citet{sri02},
indicating a significantly higher occurrence rate ($\sim$70\%).
These studies also show that
bipolar outflows from massive YSOs are much more massive and energetic
than those observed in low-mass YSOs \citep[see also][]{rm01}.
There is some controversy as to whether massive outflows are 
much less collimated than low-mass outflows (see \S~3.1).

The present paper presents \coj\ line maps of
12 massive star-forming regions 
at 27$''$ angular resolution  
(Table 1). 
Combining the results of this study with those of \citet[][2005]{zha01},
we complete a census of molecular outflows for
all sources with $l$$\ge$50$^\circ$ in
a flux-limited sample of massive protostellar candidates (see \S~2.1),
and estimate the occurrence frequency of molecular outflows 
for the sample.
We also examine the morphological, physical, and dynamical properties of 
bipolar outflows observed, and compare them with those of outflows from
low-mass YSOs.
The selection criteria of the sources and the details of the 
observations are described in \S~2. The results are 
presented in \S~3. 
The implications of our results for massive star formation 
are discussed in \S~4. 
We summarize the main results in \S~5.


\section{Observations}

\subsection{Source Selection}

Our sources are a subset of the catalog of 101 massive protostellar 
candidates \citep{mol96}, which were selected from a complete
flux-limited sample of bright IRAS point sources with far-infrared (FIR) 
colors similar
to compact molecular cores, based on the detection of NH$_3$ line emission.
The original sample consists of 260 IRAS point sources that satisfy flux,
color, and practical criteria \citep{pal91}:
1) $F_{60}$$\ge$100~Jy and no upper limits for $F_{25}$, $F_{60}$, 
and $F_{100}$, (2) 0.61$\le$log($F_{60}/F_{25}$)$\le$1.74,
and 0.087$\le$log($F_{100}/F_{60}$)$\le$0.52,
(3) no positional coincidence with known HII regions,
and (4) $|b|$$\le$10$^\circ$ and $\delta$$\ge$$-$30$^\circ$.
\citet{mol96} surveyed NH$_3$ (1,1) and (2,2) lines
towards 163 of the 260 sources, and detected 101 sources.
Among the 101 sources, 63 (``high'' group) have colors characteristic of 
UC HII regions, 
log($F_{25}/F_{12}$)$\ge$0.57 and log($F_{25}/F_{12}$)$\ge$1.30
\citep{wc89},
while the remaining 38 (``low'' group) have colors of
log($F_{25}/F_{12}$)$<$0.57. Most ($\sim$90\%) of the 101 sources have
bolometric luminosities $>$10$^3$~$L_\odot$,
so they are very likely to be massive protostars that have not yet formed 
UC HII regions, probably because of active accretion \citep[][1998]{mol96}.
\citet[][2005]{zha01} mapped 69 of the 101 sources in the CO J=2$-$1 line:
67 sources observed by \citet{mol98} in radio continuum 
and 2 other sources (\mtwo\ and \mfifteen).
They detected outflows in about 90\% (35/39) of their sources 
in the Galactic longitude range $l$$\ge$50$^\circ$.
For sources at 0$^\circ$$<$$l$$<$50$^\circ$, they could not
compile reliable statistics because outflow signatures
are frequently confused with other cloud components along
the same lines of sight.
In this study we observed 12 sources (Table 1).
They include all 11 sources with $l$$\ge$50$^\circ$ among
the 34 sources not observed by \citet{mol98}, and \mten.
Thus our sample has 3 common sources (\mtwo, \mten, and \mfifteen) with the sample
of Zhang et al. (2001, 2005).

\subsection{\coj\ Line Observations}

We mapped the 12 sources in the \coj\ line 
using the 12 m telescope\footnote{
The 12m telescope is a facility of the National Science Foundation 
currently operated by the University of Arizona Steward Observatory 
under a loan agreement with the National Radio Astronomy Observatory.
} at Kitt Peak in 2003 January and February. 
The telescope has a FWHM of 27$''$ at 230~GHz.
A 256-channel filterbank with 128~MHz bandwidth was used in parallel mode 
to observe both circular polarizations simultaneously, 
and the velocity resolution
is 0.68~\kms.
The system temperature varied in the range 500$-$700~K
during the observing sessions.
All the sources were mapped with 8$''$ spacing in position 
switching mode using the on-the-fly observing technique.
The reference positions were checked to be free from appreciable ($>$0.1~K) 
CO emission. They were usually $\sim$30$'$ away from the sources. 
The pointing and focus were checked every four hours with Venus, Saturn, 
or DR21. 
The typical $rms$ noise level is 0.2~K.
The observed temperature (\tr) was converted to main-beam
brightness temperature (\tb) using the corrected main-beam efficiency
($\eta_{\rm mb}^*$=0.51) provided by the observatory.


\section{Results}

\subsection{CO Maps}


We detected high-velocity CO gas towards all sources in our sample 
with the possible exception of \mtwoone\ (Table 1). 
Here we 
compared the observed full width at zero intensity (FWZI)
of each CO line with that expected from the FWHM by Gaussian fit, 
and define a high-velocity excess if the former is greater 
than the latter.
With the exception of one source (\monefourseven) 
we find, as do \citet{sc96a} in the CO J=1$-$0 line data, 
that the FWZI of the CO J=2$-$1 line is always $>$15~\kms\
for the sources showing high-velocity excess.
The FWZI of \monefourseven\ is about 12~\kms\ in our 0.2~K
{\it rms} data; 
possibly it would be broader at the 0.02~K {\it rms} level of
\citet{sc96a}.
 
Based on contour maps of blue- and red-shifted high-velocity gas and 
position-velocity diagrams, we conclude that 10 of the 11 sources 
host outflows; the exception is \msixteen\ (Fig. 1).
In \monefourseven\ only blue-shifted high-velocity gas is detected.
This is consistent with the result of CO J=1$-$0 line observations
made by \citet{sug89} at 17$''$ resolution. 
The IRAS sources are located reasonably close to the outflow centers
in most cases, but they are offset from the centers by
greater than the 27$''$ telescope beamwidth in three sources (\mnineteen, \mtwoseven, \monesixthree). 
In these cases the IRAS sources are unlikely to be the driving sources 
of the outflows. 
Some maser sources and/or (sub)millimeter continuum sources, which are
good indicators of massive star-forming sites in early evolutionary
stages,
are found around the outflow centers in \mtwoseven\ \citep{wol03} and 
\monesixthree\ (Beuther et al. 2002b; see also Anglada \& Rodr\'\i{}guez 2002).

%
The morphologies of blue- and red-lobes are roughly symmetric in
five sources (\mtwo, \mten, \mfifteen, \monetwofour, \monefoursix),
whereas they are asymmetric in the remaining 4 sources (\mfourteen, 
\mnineteen, \mtwoseven, \monesixthree).
Such an asymmetry is often seen in the CO maps of outflows both from
low-mass and high-mass YSOs 
\citep[e.g.,][]{bon96, beu02a, zha05}.
Possible explanations of this feature are
the asymmetric ejection of outflow-driving agents 
from the central stars,
the nonuniform distribution of the ambient molecular gas,
and the presence of multiple outflows.
There seem to be two (or more) outflows in \mfifteen.
This may be true for \mten\ as well \citep[see also][]{hun95}.
This multiplicity complicates
the high-velocity gas distributions of the two regions.
We estimate the physical and dynamical parameters of 
the two (A and B) outflows in \mfifteen\ separately (see \S~3.2),
while we do not for the two of \mten, because they are entangled
in position as well as in velocity.

%
As noted earlier, there is some debate about the collimation of 
outflows from massive YSOs.
Richer et al. (2000) and Ridge \& Moore (2001) argued 
that outflows from massive YSOs are significantly less collimated 
than those observed in low-mass YSOs (see also Wu et al. 2004 and \S~4),
while Beuther et al. (2002a) suggested, 
based on higher-resolution (11$''$) CO maps, that they 
are better collimated than previously thought.
We determine the collimation factors ($f_{\rm c}$), 
defined as the ratio of outflow length to width, 
of the 11 observed outflows (Table 2).
The values range between 1.0 and 3.8.
They are greater than 2.0 for 6 sources; 
for comparison, Ridge \& Moore (2001) reported collimation factors 
of $\sim$1$-$2.
Our average collimation factor is 2.3, 
similar to the 2.1 value determined by \citet{beu02a}
for 15 outflows from massive protostellar candidates 
despite our factor of $\sim$2.5 lower angular resolution.
This may be because all outflows in our sample
are located closer to the Sun 
and in less crowded (75$^\circ$$<$$l$$<$205$^\circ$) regions
of the Galactic plane than the majority of their sources.
The average value is also very similar to that (2.2) measured for 
36 low-mass outflows as discussed in \S~4.
From our sample, \mten~A, \mfourteen, and \mtwoseven\ are particularly 
well collimated with factors $>$3.0.
In comparison, none of outflows observed by \citet{beu02a} 
have collimation factors greater than 3.0.

\subsection{Outflow Parameters}

We derive the physical and dynamical parameters of the outflows
in a similar way to \citet{cb92} and \citet{beu02a}.
Here we assume (1) the excitation temperature of CO gas is 30~K 
\citep{mol96},
(2) the integrated intensity ratio of CO to $^{13}$CO is 10 in the
line-wing velocity range \citep{cho93},
(3) [CO]/[H$_2$]=1$\times$10$^{-4}$ and [CO]/[$^{13}$CO]=89,
and (4) the mean molecular weight is 2.3.
We do not correct for the inclination of the outflow.
Table 2 displays the results: masses $M_{\rm b}$ (blue-shifted),
$M_{\rm r}$ (red-shifted), $M_{\rm out}$ (total), momentum $P$, 
kinetic energy $E$, dynamical age $t$, mass outflow rate ${\dot{M}_{\rm out}}$,
mechanical force $F_{\rm m}$, mechanical luminosity $L_{\rm m}$,
and collimation factor $f_{\rm c}$.
The vast majority of outflows have 
sizes of 0.3$-$1.4~pc, dynamical ages of (1$-$10)$\times$10$^4$~yr,
masses $>$10~\msol, momenta $>$100~\msol~\kms, 
and mass outflow rates of (1$-$10)$\times$10$^{-4}$~\msol~yr$^{-1}$.
Thus these outflows are much more massive and energetic
than outflows from low-mass protostars (Class 0 and 
Class I objects) with
similar dynamical ages \citep[e.g.,][]{cb92, bon96, wu04}. 
This is consistent with the results of previous
single-dish studies of massive YSO outflows
\citep[][]{sc96b, beu02a, zha05}. 
The large estimated outflow rates suggest mass accretion rates 
large enough to overcome the strong radiation of the central
massive stars (see \S~4 for details).

\subsection{Occurrence Frequency of Outflows}

%
\citet[][2005]{zha01} mapped 69 of the 101 sources of 
\citet{mol96} in the \coj\ line at about 30$''$ grid spacing
using the Kitt Peak 12~m and the Caltech Submillimeter Observatory 10~m.
They detected outflows in 39 of the 65 sources that have data suitable
for outflow identification, 
for a detection rate of 60\%. 
But it was practically impossible to compile reliable statistics 
for 26 sources in the Galactic longitude range 0$^\circ$$<$$l$$<$50$^\circ$
because of multiple cloud components in the same lines of sight.
They found a much higher 90\% (35/39) detection rate for 39 sources 
with $l$$\ge$50$^\circ$.
In this study we observed 12 sources selected from the same catalog.
They are all located at $l$$\ge$50$^\circ$.
We detected in 10 sources outflows that have similar physical and
dynamical properties to other outflows from massive YSOs.
This indicates the outflow detection rate is 88\% for all 48 sources
with $l$$\ge$50$^\circ$ in the Molinari catalog,
taking into account that the two samples have 3 sources in common.
It may be reasonable to expect a similar occurrence frequency of 
outflows for the sources at 0$^\circ$$<$$l$$<$50$^\circ$,
because both groups were selected by the same criteria.
The FWZI of the CO line is $\sim$10~\kms\ for the sources that show
no distinct high-velocity wings. 
Assuming that the typical outflow velocity is $\sim$25~\kms,
outflows at inclinations $\gtrsim$80$^\circ$ would not be
detected. Statistically, the fraction of those outflows is
$\sim$10\%.
Thus this result strongly suggests that nearly all sources in the
Molinari catalog have molecular outflows.

%
There have been several attempts to measure the occurrence frequency of 
molecular outflows in low-mass protostellar objects 
\citep[e.g.,][]{ter89, par91, bon96}.
The detection rates range between $\sim$70\% and 90\%.
For example, \citet{bon96} mapped 
45 low-mass protostars (36 Class I and 9 Class 0 objects) 
of nearby ($<$450~pc) molecular clouds in the \coj\ line,
and found outflows in 75\%$-$80\% of the sources.
The detection rate was higher (80\%$-$90\%) for 34 YSOs with
dust continuum emission. 
These values are similar to the detection rate determined by 
this study for the Molinari sources.
Therefore, bipolar outflows appear to be 
an integral stage in the formation process of massive stars as well,
as suggested by some previous studies
(e.g., Zhang et al. 2001, 2005; Beuther et al. 2002a; Wu et al. 2004).


\section{Implications for Massive Star Formation}

%
As mentioned earlier, the formation process of massive stars
is still poorly understood. Two major theoretical models are 
currently competing: accretion and coalescence.
If massive stars form by accretion as do low-mass stars, 
all massive protostars would generate massive and powerful outflows.
Probably this is not true for the coalescence model
where massive stars form from the collision of low-mass (proto)stars.
There are three major collision units, 
such as stars, disks(/outflows), and cores, 
and six types of collisions are possible among the three \citep{bz05}.
Only if disk-disk collisions dominate the other types of collisions,
and if the collisions do not profoundly disrupt the outflows
of low-mass (proto)stars,
might one expect that most massive protostars have massive 
and energetic outflows.
Thus the observed high occurrence frequency of outflows
can not be easily explained by the coalescence model.
On the other hand,
the collimation degree of outflows could be very different
between the two models.
The masses of massive outflows are usually one order of magnitude larger
than those ($<$1~\msol) of low-mass outflows.
If massive stars form by coalescence,
this suggests that a massive outflow may arise after multiple collisions. 
In that case,
although the outflows of low-mass protostars survive the collisions,
the resulting massive outflow might be highly impulsive and 
poorly collimated like the OMC-1 outflow observed by \citet{all93}.
In contrast, one would expect relatively well-collimated outflows 
around massive protostars in the accretion model.

Most outflows observed by this study do {\it not} seem to be 
very different from many low-mass outflows in collimation 
\citep[e.g.,][]{bon96}. 
In order to investigate this quantitatively,
we compare the frequency distribution of collimation factors of 
40 massive outflows
with that of 36 low-mass outflows.
In this comparison we use the values derived by \citet{lad85}, 
\citet{beu02a},
and this study for 14, 15, and 11 massive outflows, respectively.
The associated IRAS sources almost always have bolometric luminosities 
from 10$^3$ to 10$^5$~\lsol, and the majority of them are massive protostellar
candidates.
For low-mass outflows
we adopt the values determined by \citet{lad85} for 12 sources,
and estimate collimation factors for 24 sources 
using the CO maps provided by
\citet{mye88}, \citet{sch88}, \citet{par91}, and \citet{bon96},
which are systematic single-dish studies of multiple sources
mainly at relatively low (30$''$$-$45$''$) resolutions.
Here we exclude CO maps that do not cover the full outflows.
Most of their driving sources are Class 0 and Class I objects. 
Figure 2 shows the result.
Though a higher fraction of low-mass outflows 
have collimation factors $\ge$3.5, 
the two distributions appear to be quite similar and
the average values, 2.2, are the same.
This does not agree with the finding of \citet{wu04} 
for a sample of 231 YSOs that
the average collimation factors are 2.8$\pm$2.2 and 2.1$\pm$1.0 
for low-mass and massive outflows, respectively.
This discrepancy may be mostly because of different resolutions of 
the data used for low-mass outflows in the analyses,
even if we cannot exclude the possibility that some intrinsic 
difference exists between low-mass and massive outflow collimations.
Our analysis includes only single-dish data for all low-mass outflows, 
while the Wu's analysis contains significantly higher-resolution single-dish 
and interferometer data for many low-mass outflows,
which can reveal highly-collimated outflows.
There are 10 outflows with collimation factors $>$5 in
the Wu sample of low-mass outflows.
Unless they are included, the average collimation factor 
is close to 2.2.
On the other hand,
it is not straightforward to identify highly-collimated outflows
around massive YSOs even in interferometer data,
because massive stars tend to form in groups and clusters 
where multiple outflows may overlap. 
For example, Beuther et al. (2002c)
found at least 3 outflows, two of which are highly collimated,
in the interferometer data
of the massive star-forming region 05358+3543. 
The three could not be separated in single-dish data
and the collimation factor of the combined outflow was
measured to be 2.0 (Beuther et al. 2002a). 

It is widely accepted that low-mass outflows are momentum-driven 
by highly-collimated jets and/or wide-angle winds from the central
(proto)stars (K$\ddot{\rm o}$nigl \& Pudritz 2000; Shu et al. 2000).
If massive outflows are produced by the same mechanisms,
the mass accretion rates can be estimated from the observed outflow rates
under the assumption of momentum conservation between jets/winds and
outflows. 
The ratio of wind mass-loss rate to accretion rate is typically $\sim$0.1
in the highly-collimated jet model \citep{kp00} 
and $\sim$0.3 in the wide-angle wind model \citep{shu00}. 
Assuming that the velocity ratio of winds to outflows is around 20 
\citep[see \S~4.2 of][]{beu02a},
the mass accretion rate would be 20\%$-$50\% of the outflow rate.
The average value of the observed outflow rates
is 5$\times$10$^{-4}$~$M_\odot$~yr$^{-1}$ (Table 2),
so the accretion rate is $\sim$1$\times$10$^{-4}$~$M_\odot$~yr$^{-1}$.
This value is much greater 
than the typical accretion rates of (10$^{-7}$$-$10$^{-5}$)~$M_\odot$~yr$^{-1}$
expected and measured in low-mass YSOs 
\citep[e.g.,][]{shu77, bon96}.
It is also large enough to overcome the radiation pressure
of the central stars, which have bolometric luminosities 
$\le$2$\times$10$^4$~\lsol\ \citep{wal95}.
Therefore, the results of this study appear to support the accretion
model rather than the coalescence model.

\section{Conclusions}

We mapped 12 massive protostellar candidates in the \coj\ line,
which in combination with \citet{zha05}
completed a census of bipolar molecular outflows 
for all 48 sources with $l$$>$50$^\circ$ 
in the Molinari catalog of 101 massive protostellar candidates.
We detected outflows in 10 of the 12 sources,
indicating 88\% occurrence frequency of outflows in the 48 sources.
The observed outflows are much more massive ($>$10~\msol)
and energetic ($>$100~\msol~\kms) than those observed in low-mass YSOs.
These results 
imply that almost all sources of the Molinari catalog 
have very massive and powerful molecular outflows.
Thus bipolar outflows appear to be a ubiquitous phenomenon 
in massive star formation as in low-mass star formation,
as suggested by some previous studies.

Nearly all outflows observed by this study have mass outflow rates
$>$2$\times$10$^{-4}$~\msol~yr$^{-1}$,
suggesting large accretion rates of
$\sim$1$\times$10$^{-4}$~$M_\odot$~yr$^{-1}$.
These values are much greater than the typical accretion rates
expected and measured in low-mass YSOs and are large enough to overcome
the radiation pressure of the central massive protostars.
We compared the collimation factors of 40 massive outflows
with those of 36 low-mass outflows, 
and found {\it no} significant difference between them.
This can be naturally understood in the accretion model, 
but cannot be easily explained by the coalescence model. 
Therefore, it is likely that accretion plays a dominant role
in the formation process of massive stars, although we cannot exclude
the possibility that coalescence plays a role in some special circumstances.

We acknowledge support from the Laboratory for Astronomical Imaging at the 
University of Illinois and NSF AST 0228953.


\clearpage


\begin{deluxetable}{cccccccccccccc}
\rotate
\tabletypesize{\footnotesize}
\tablewidth{0pt}
\tablecaption{Source Summary}
\tablehead
{
\colhead{Source\tablenotemark{a}} & \colhead{IRAS} & & 
\colhead{$\alpha$(J2000)} & \colhead{$\delta$(J2000)} & 
\colhead{$l$} & \colhead{$b$} & \colhead{$d$} & \colhead{$L_{\rm bol}$} &
\multicolumn{2}{c}{Maser} & \colhead{} & \multicolumn{2}{c}{\coj} 
                 \\ \cline{10-11} \cline{13-14}
\colhead{Number} & \colhead{Name} & \colhead{Type\tablenotemark{a}} &
\colhead{($^{\rm h}~ ^{\rm m}~ ^{\rm s}$)} & \colhead{($^\circ~ '~ ''$)} & 
\colhead{($^\circ$)} & \colhead{($^\circ$)} &
\colhead{(kpc)} & \colhead{($L_\odot$)} & \colhead{H$_2$O\tablenotemark{b}} &
\colhead{CH$_3$OH\tablenotemark{c}} & \colhead{} & \colhead{Wings} & \colhead{Outflow}
}
\startdata
~~2 & 00117+6412~~ & H & 00 14 27.7 & +64 28 46 & 118.961  & +1.893 & 
	1.8 & 1.38E3 &  Y & N & & Y & Y \\ 
~10 & 05274+3345~~ & H & 05 30 45.6 & +33 47 52 & 174.197 &  $-$0.078 &
	1.6 & 4.35E3 &  Y & Y & & Y & Y \\
~14 & 05553+1631~~ & H & 05 58 13.9 & +16 32 00 & 192.161 &  $-$3.815 &
	2.0\tablenotemark{d} & 5.20E3 &  Y & N & & Y & Y \\
~15 & 06056+2131~~ & H & 06 08 41.0 & +21 31 01 & 189.032 &  +0.784 &
	2.0\tablenotemark{d} & 1.04E4 &  Y & Y & & Y & Y \\
~16 & 06061+2151~~ & H & 06 09 07.8 & +21 50 39 & 188.796 &  +1.033 &
	2.0\tablenotemark{d} & 1.11E4 &  Y & Y & & Y & N \\
~19 & 06103+1523~~ & H & 06 13 15.1 & +15 22 36 & 194.934 &  $-$1.227 &
	4.6 & 1.91E4 &  N & N & & Y & Y \\
~21 & 06105+1756~~ & H & 06 13 28.3 & +17 55 30 & 192.723 &  +0.040 &
	3.4 & 1.60E4 &  N & N & & N? & N \\
~27 & 06382+0939~~ & H & 06 41 02.7 & +09 36 10 & 203.205 &  +2.080 &
	0.8 & 1.63E2 & N & N & & Y & Y \\
124 & 20227+4154~~ & L & 20 24 31.4 & +42 04 17 & ~79.885 &   +2.552 &
	1.7\tablenotemark{e} & 2.64E3 &  Y & N & & Y & Y \\
146 & 22267+6244~~ & H & 22 28 29.4 & +62 59 44 & 107.504 &  +4.488 &
	0.5 & 1.10E2 & N & N & & Y & Y \\
147 & 22272+6358A & H & 22 28 52.3 & +64 13 43 & 108.186 &   +5.519 &
	1.2 & 1.97E3 & N & Y & & Y & Y \\
163 & 23545+6508~~ & H & 23 57 05.2 & +65 25 11 & 117.315 &  +3.142 &
	1.3 & 3.89E3 & N & N & & Y & Y \\
\enddata
\tablenotetext{a}{Source number and High/Low (H/L) designation of \citet{mol96}.}
\tablenotetext{b}{\citet{wou93} and \citet{bra94}.}
\tablenotetext{c}{\citet{szy00}.}
\tablenotetext{d}{Assumed to be located in the Gemini OB1 cloud complex 
\citep{car95}, in which case the $L_{\rm bol}$'s were re-calculated 
with new distances from the values given by \citet{mol96}.}
\tablenotetext{e}{Assumed to be associated with the Cygnus~X complex \citep{dt85}.}
\end{deluxetable}

\clearpage


\begin{deluxetable}{lccccccccccc}
\rotate
\tabletypesize{\footnotesize}
\tablewidth{0pt}
\tablecaption{Outflow Properties Derived from \coj\ Line Data}
\tablehead
{
\colhead{} & \colhead{} & \colhead{} & \colhead{} & \colhead{} & 
\colhead{} & \colhead{$E$} & \colhead{$t$} &
\colhead{${\dot{M}_{\rm out}}$} & \colhead{$F_{\rm m}$} & 
\colhead{} & \colhead{}\\
\colhead{IRAS} & \colhead{Size} & \colhead{$M_{\rm b}$} & 
\colhead{$M_{\rm r}$} & \colhead{$M_{\rm out}$} & 
\colhead{$P$} & \colhead{(10$^{46}$)} & \colhead{(10$^{4}$)} &
\colhead{(10$^{-4}$)} & \colhead{(10$^{-3}$)} & 
\colhead{$L_{\rm m}$} & \colhead{}\\
\colhead{Name} & \colhead{(pc)} & \colhead{($M_\odot$)} &
\colhead{($M_\odot$)} & \colhead{($M_\odot$)} &
\colhead{($M_\odot$~\kms)} & \colhead{(10$^{46}$~erg)} & 
\colhead{(10$^4$~yr)} & \colhead{($M_\odot$~yr$^{-1}$)} & 
\colhead{($M_\odot$~\kms~yr$^{-1}$)} & \colhead{($L_\odot$)} &
\colhead{$f_{\rm c}$}
}
\startdata
00117+6412 &    0.7 &	6.6 &  5.6 &   12.1 &  213 &   3.7 &   3.8 &   3.2 &   5.6 &   7.8 & 1.0 \\
05274+3345A/B\tablenotemark{a} 
    &   1.4 &	20.1~ & 14.7~ &  34.8 &  481 &   6.7 &   9.9 &   3.5 &   4.9 &   5.4 & 3.3/1.6\\
05553+1631~~ &    1.7 &	29.0~ &  8.9 &   37.9 &  609 &   9.9 &  11.1~ &   3.4 &   5.5 &   7.1 & 3.4 \\
06056+2131A &   1.4 & 	21.7~ &   29.4~ &  51.1 &  694 &   9.4 &  10.1~ &   5.1 &   6.9 &   7.4 & 2.0 \\
06056+2131B &    1.0 & 35.5~ &   7.7 &   43.2 &  617 &   8.9 &   7.5 &   5.8 &   8.3 &   9.5 & 1.0 \\
06103+1523 &   2.2 & 28.0~ &   37.1~ &   65.1 &  995 &  15.4~ &  14.4~ &   4.5 &   6.9 &   8.5 & 2.7 \\
06382+0939 &    0.9 &	4.3 &   8.9 &    13.2 &  193 &   2.9 &   5.8 &   2.3 &   3.3 &   4.0 & 3.8 \\
20227+4154 &   1.2 & 56.0~ &  34.7~ &   90.6 & 2937~ & 98.3~ &   3.8 &  23.8~ &  77.3~ & 205.1~ & 2.1 \\
22267+6244 &	0.3 &	2.0 &    1.8 &    ~3.8 &  ~66 &   1.2 &   1.6 &   2.4 &   4.2 &   5.9 & 1.5 \\
22272+6358A\tablenotemark{b} &	0.4 &	1.4 &    0.0 &    ~~1.4 &   ~16 &   0.2 &   3.4 &   0.4 &   0.5 &   0.4 & \nodata \\
23545+6508 &    0.8 &	3.5 &    2.1 &    ~5.5 &   ~61 &   0.7 &   7.2 &   0.8 &   0.9 &   0.7 & 2.8 \\
\enddata
\tablenotetext{a}{The properties of 05274+3345A and 05274+3345B 
are not separately estimated except for the collimation factor, 
because the two are entangled both in position and velocity.}
\tablenotetext{b}{Only blue-shifted high-velocity gas is detected
(see the text).}
\end{deluxetable}

\clearpage



\begin{figure}[tbp]
\vskip 0cm
\figurenum{1}
\epsscale{0.9}
\plotone{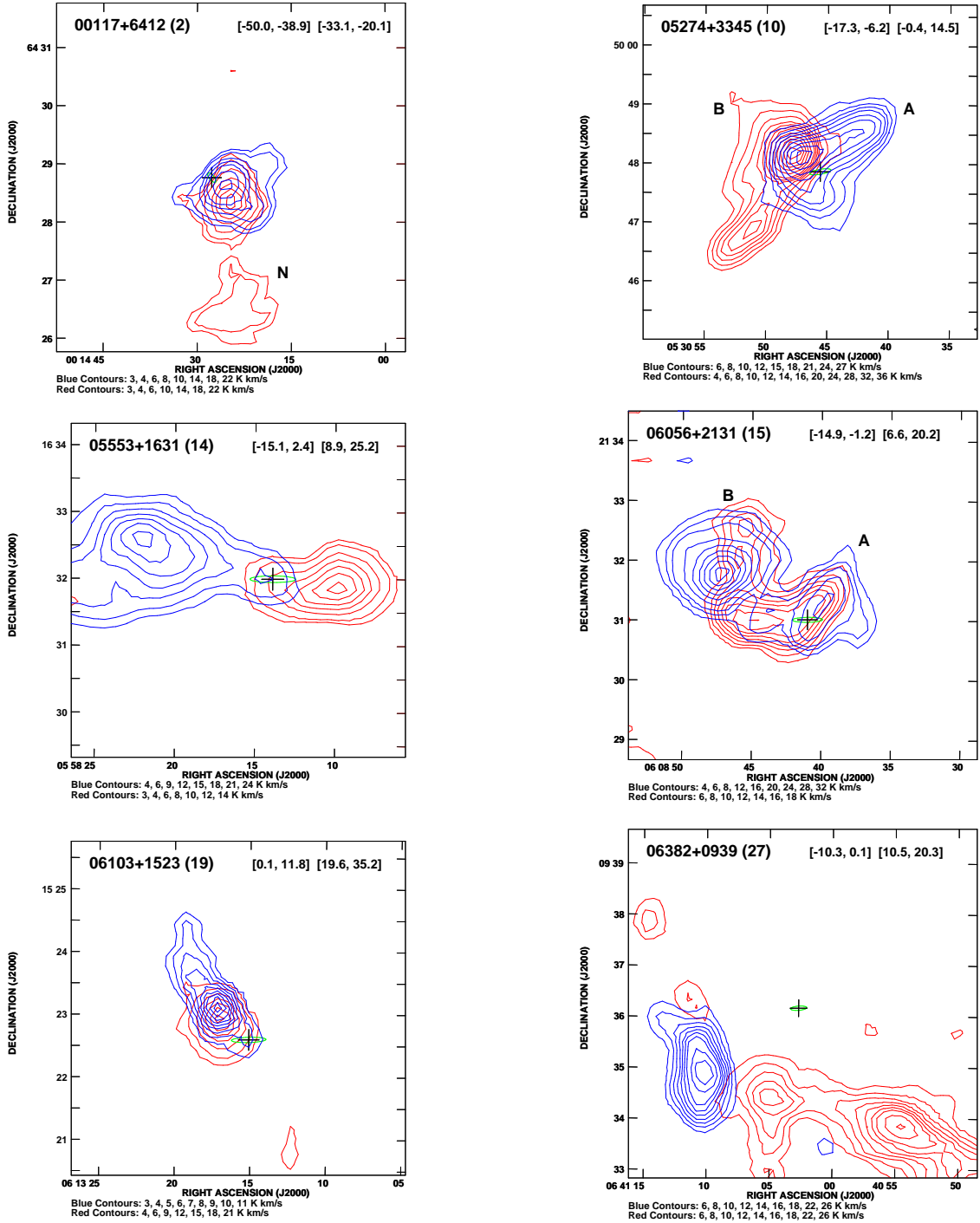}
\vskip 0cm
\figcaption{\footnotesize
\coj\ line integrated intensity maps. 
The Molinari source designation is given in parentheses following the
IRAS name.
Blue contours represent blue-shifted
high-velocity gas, while red contours indicate red-shifted gas. 
The integrated velocity ranges are presented at the top right corner 
of each panel.
Contour levels are listed at the bottom in each panel.
The cross and its associated ellipse show the position of 
the IRAS point source and its uncertainty, respectively.
The first and last panels show features marked by ``N'' which were not
included when deriving the outflow parameters. In the first panel
(\mtwo) the ``N'' contours are a separate velocity component,
whereas in the last panel (\monesixthree) the contribution of the ``N''
component is not well-determined because it is not fully
imaged.
}
\end{figure}

\begin{figure}[tbp]
\vskip 0cm
\figurenum{1}
\epsscale{0.9}
\plotone{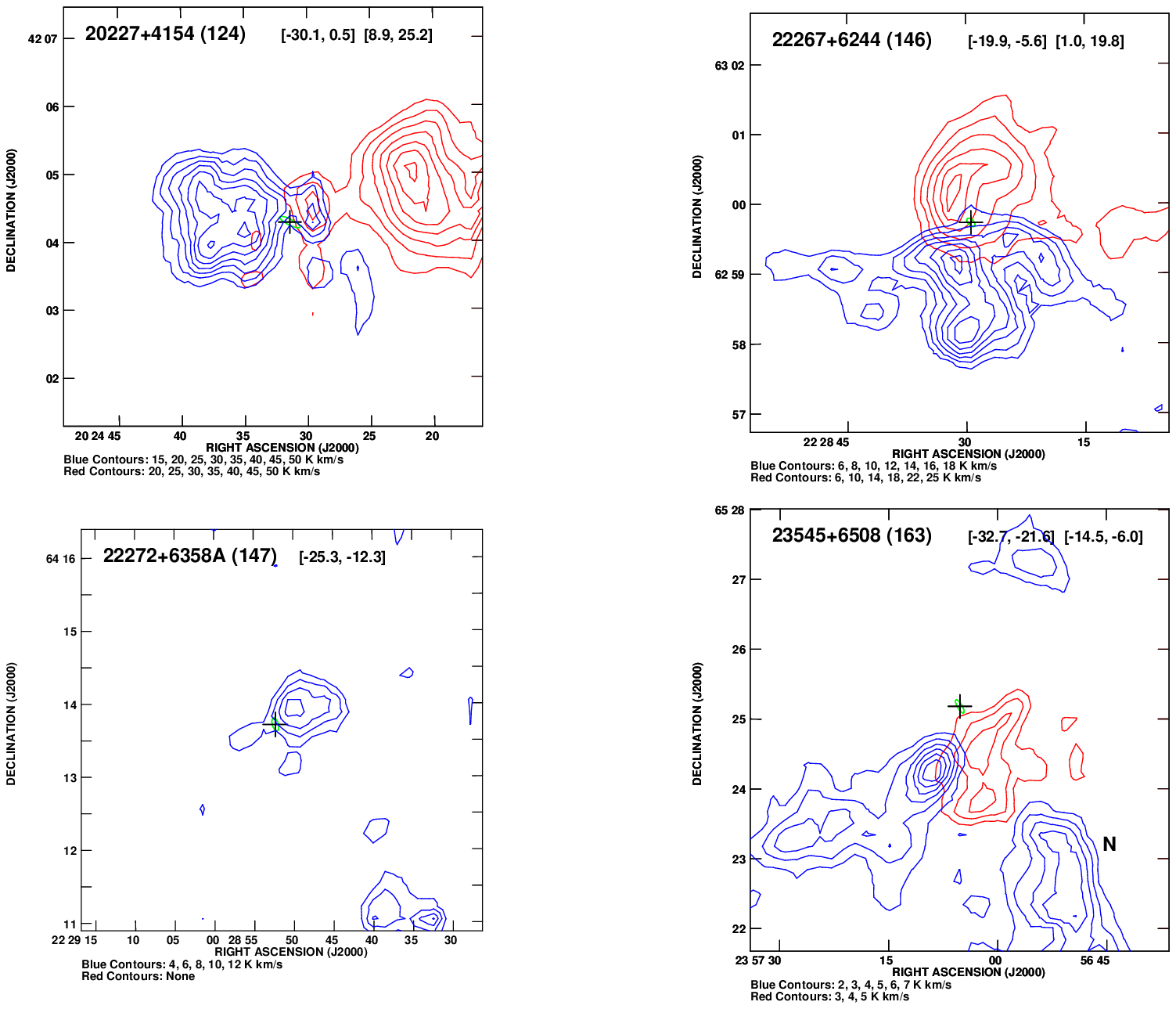}
\vskip 0cm
\figcaption{{\it Continued}
}
\end{figure}


\begin{figure}[tbp]
\vskip 0cm
\figurenum{2}
\epsscale{1.0}
\plotone{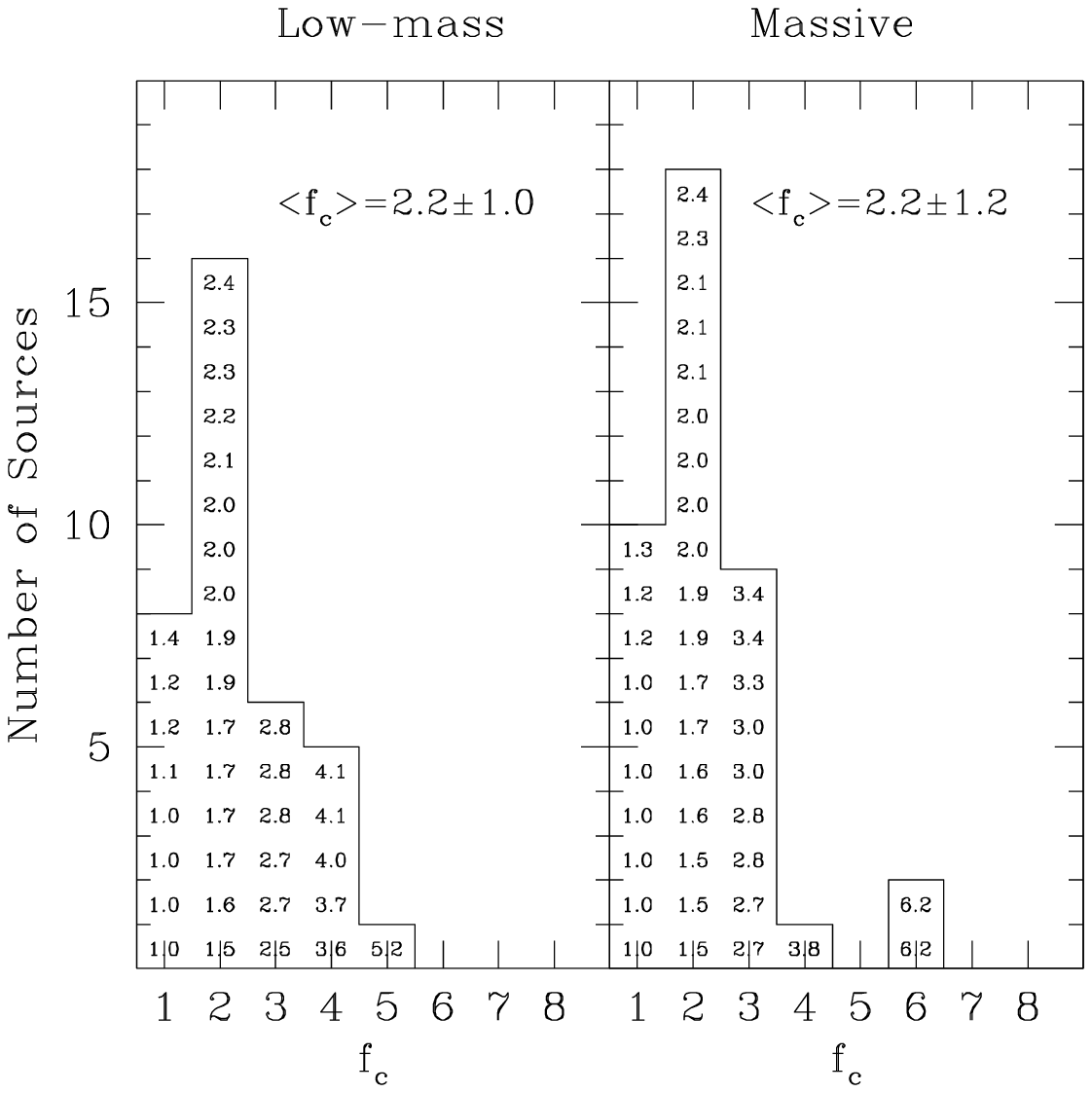}
\vskip 0cm
\figcaption{
Frequency distributions of collimation factors for 
({\it left}) 36 low-mass outflows and ({\it right}) 40 massive outflows. 
The values are listed in each bin. The two groups have the same average
value, 2.2.
}
\end{figure}


\begin{thebibliography}{}

\bibitem[Allen \& Burton(1993)]{all93} Allen, D. A., \& Burton, M. G. 1993, Nature, 363, 54 
\bibitem[Anglada \& Rodr\'\i{}guez(2002)]{ang02} Anglada, G., \& Rodr\'\i{}guez, L. F. 2002, RMxA\&A, 38, 13
\bibitem[Bachiller(1996)]{bac96} Bachiller, R. 1996, \araa, 34, 111
\bibitem[Bally \& Zinnecker(2005)]{bz05} Bally, J., \& Zinnecker, H. 2005, AJ, 129, 2281 
\bibitem[Beuther et al.(2002c)]{beu02b} Beuther, H., Schilke, P., Gueth, F., McCaughrean, M., Andersen, M., Sridharan, T. K., \& Menten, K. M. 2002c, \aap, 387, 931 
\bibitem[Beuther et al.(2002b)]{beu02b} Beuther, H., Schilke, P., Menten, K. M., Motte, F., Sridharan, T. K., \& Wyrowski, F. 2002b, \apj, 566, 945
\bibitem[Beuther et al.(2002a)]{beu02a} Beuther, H., Schilke, P., Sridharan, T. K., Menten, K. M., Walmsley, C. M., \& Wyrowski, F. 2002a, \aap, 383, 892 
\bibitem[Bonnell et al.(1998)]{bon98} Bonnell, I. A., Bate, M. R., \& Zinnecker, H. 1998, MNRAS, 298, 93
\bibitem[Bontemps et al.(1996)]{bon96} Bontemps, S., Andre, P., Terebey, S., \& Cabrit, S. 1996, \aap, 311, 858
\bibitem[Brand et al.(1994)]{bra94} Brand, J. et al. 1994, \aaps, 103, 541 
\bibitem[Cabrit \& Bertout(1992)]{cb92} Cabrit, S., \& Bertout, C. 1992, \aap, 261, 274
\bibitem[Carpenter et al.(1995)]{car95} Carpenter, J. M., Snell, R. L., \& Schloerb, F. P. 1995, \apj, 450, 201 
\bibitem[Choi et al.(1993)]{cho93} Choi, M., Evans, N. J. II, \& Jaffe, D. T. 1993, \apj, 417, 624
\bibitem[Dame \& Thaddeus(1985)]{dt85} Dame, T. M., \& Thaddeus, P. 1985, \apj, 297, 751 
\bibitem[Hunter et al.(1995)]{hun95} Hunter, T. R., Testi, L., Taylor, G. B., Tofani, G., Felli, M., \& Phillips, T. G. 1995, A\&A, 302, 249
\bibitem[K$\ddot{\rm o}$nigl \& Pudritz (2000)]{kp00} K$\ddot{\rm o}$nigl, A. \& Pudritz, R. E. 2000, in Protostars and Planets IV, ed. V. Mannings, A. P. Boss, \& S. S. Russell (Tucson: Univ. Arizona Press), 759 
\bibitem[Lada(1985)]{lad85} Lada, C. J. 1985, ARA\&A, 23, 267 
\bibitem[McKee \& Tan(2003)]{mt03} McKee, C. F., \& Tan, J. C. 2003, \apj, 585, 850
\bibitem[Molinari et al.(1996)]{mol96} Molinari, S., Brand, J., Cesaroni, R., \& Palla, F. 1996, \aap, 308, 573
\bibitem[Molinari et al.(1998)]{mol98} Molinari, S., Brand, J., Cesaroni, R., Palla, F., \& Palumbo, G. G. C. 1998, \aap, 336, 339 
\bibitem[Myers et al.(1988)]{mye88} Myers, P. C., Heyer, M., Snell, R. L., \& Goldsmith, P. F. 1988, \apj, 324, 907 
\bibitem[Palla et al.(1991)]{pal91} Palla, F., Brand, J., Comoretto, G., Felli, M., \& Cesaroni, R.  1991, \aap, 246, 249
\bibitem[Parker et al.(1991)]{par91} Parker, N. D., Padman, R., \& Scott, P. F. 1991, \mnras, 252, 442
\bibitem[Ridge \& Moore(2001)]{rm01} Ridge, N. A., \& Moore, T. J. T. 2001, \mnras, 378, 495
\bibitem[Schwartz et al.(1988)]{sch88} Schwartz, P. R., Gee, G., \& Huang, Y.-L. 1988, \apj, 327, 350 
\bibitem[Shepherd \& Churchwell(1996a)]{sc96a} Shepherd, D. S., \& Churchwell, E. 1996a, \apj, 457, 267
\bibitem[Shepherd \& Churchwell(1996b)]{sc96b} Shepherd, D. S., \& Churchwell, E. 1996b, \apj, 472, 225
\bibitem[Shu (1977)]{shu77} Shu, F. H. 1977, \apj, 214, 488 
\bibitem[Shu et al.(1987)]{shu87} Shu, F. H., Adams, F. C., \& Lizano, S. 1987, ARA\&A, 25, 23 
\bibitem[Shu et al.(2000)]{shu00} Shu, F. H., Najita, J. R., Shang, H., \& Li, Z.-Y. 2000, in Protostars and Planets IV, ed. V. Mannings, A. P. Boss, \& S. S. Russell (Tucson: Univ. Arizona Press), 789 
\bibitem[Sridharan et al.(2002)]{sri02} Sridharan, T. K., Beuther, H., Schilke, P., Menten, K. M., \& Wyrowski, F. 2002, \apj, 566, 931 
\bibitem[Sugitani et al.(1989)]{sug89} Sugitani, K., Fukui, Y., Mizuni, A., \& Ohashi, N. 1989, \apj, 342, L87 
\bibitem[Szymczak et al.(2000)]{szy00} Szymczak, M., Hrynek, G., \& Kus, A. J. 2000, \aaps, 143, 269 
\bibitem[Terebey et al.(1989)]{ter89} Terebey, S., Vogel, S. N., \& Myers, P. C. 1989, \apj, 340, 472
\bibitem[Walmsley(1995)]{wal95} Walmsley, M. 1995, RMxAC, 1, 137 
\bibitem[Wolf-Chase et al.(2003)]{wol03} Wolf-Chase, G., Moriarty-Schieven, G., Fich, M., \& Barsony, M. 2003, \mnras, 344, 809
\bibitem[Wolfire \& Cassinelli(1987)]{wc87} Wolfire, M. G., \& Cassinelli, J. P. 1987, \apj, 319, 850 
\bibitem[Wood \& Churchwell(1989)]{wc89} Wood, D. O. S., \& Churchwell, E. 1989, \apj, 340, 265
\bibitem[Wouterloot et al.(2004)]{wou93} Wouterloot, J. G. A., Brand, J., \& Fiegle, K. 1993, \aaps, 98, 589 
\bibitem[Wu et al.(2004)]{wu04} Wu, Y., Wei, Y., Zhao, M., Shi, Y., Yu, W., Qin, S., \& Huang, M. 2004, \aap, 426, 503 
\bibitem[Yorke \& Sonnhalter(2002)]{ys02} Yorke, H. W., \& Sonnhalter, C. 2002, \apj, 569, 846
\bibitem[Zhang et al.(2001)]{zha01} Zhang, Q., Hunter, T. R., Brand, J., Sridharan, T. K., Molinari, S., Kramer, M. A., \& Cesaroni, R.  2001, \apj, 552, L167 
\bibitem[Zhang et al.(2005)]{zha05} Zhang, Q., Hunter, T. R., Brand, J., Sridharan, T. K., Cesaroni, R., Molinari, S., Wang, J., \& Kramer, M. A. 2005, \apj, 625, 864 

\end{thebibliography}
\end{document}